# The AIV strategy of the Common Path of Son of X-Shooter

Federico Biondi[a,b], Kalyan Kumar Radhakrishnan Santhakumari[a], Riccardo Claudi[a], Matteo Aliverti[c], Luca Marafatto[a], Davide Greggio[a], Marco Dima[a], Gabriele Umbriaco[a,d], Nancy Elias-Rosa[a], Sergio Campana[c], Pietro Schipani[e], Andrea Baruffolo[a], Sagi Ben-Ami[f], Giulio Capasso[e], Rosario Cosentino[g], Francesco D'Alessio[h], Paolo D'Avanzo[c], Ofir Hershko[f], Hanindyo Kuncarayakti[i], Marco Landoni[c], Matteo Munari[j], Giuliano Pignata[k], Adam Rubin[l], Salvatore Scuderi[j], Fabrizio Vitali[h], David Young[m], Jani Achrén[n], José Antonio Araiza-Durán[o], Iair Arcavi[p], Anna Brucalassi[k,q], Rachel Bruch[f], Enrico Cappellaro[a], Mirko Colapietro[e], Massimo Della Valle[e], Marco De Pascale[a], Rosario Di Benedetto[j], Sergio D'Orsi[e], Avishay Gal-Yam[f], Matteo Genoni[c], Marcos Hernandez Diaz[g], Jari Kotilainen[i], Gianluca Li Causi[r], Seppo Mattila[s], Michael Rappaport[f], Davide Ricci[a], Marco Riva[c], Bernardo Salasnich[a], Stephen Smartt[m], Ricardo Zánmar Sánchez[j], Maximilian Stritzinger[t], Héctor Pérez Ventura[g]

[a]INAF - Osservatorio Astronomico di Padova, Vicolo dell'Osservatorio 5, I-35122 Padova, Italy;
[b]Max-Planck-Institut für Extraterrestrische Physik, Giessenbachstr. 1, D-85748 Garching, Germany;
[c]INAF – Osservatorio Astronomico di Brera, Via Bianchi 46, I-23807 Merate (LC), Italy;
[d]Università degli Studi di Padova, Vicolo dell'Osservatorio 3, I-35122 Padova, Italy;
[e]INAF – Osservatorio Astronomico di Capodimonte, Salita Moiariello 16, I-80131 Napoli, Italy;
[f]Weizmann Institute of Science, Herzl St 234, Rehovot, 7610001, Israel;
[g]INAF - Fundación Galileo Galilei, Rambla J.A. Fernández Pérez 7, E-38712 Breña Baja (TF), Spain;
[h]INAF – Osservatorio Astronomico di Roma, Via Frascati 33, I-00078 Monte Porzio Catone, Italy;
[i]FINCA - Finnish Centre for Astronomy with ESO, FI-20014 University of Turku, Finland;
[j]INAF – Osservatorio Astronomico di Catania, Via S. Sofia 78, I-95123 Catania, Italy;
[k]Universidad Andrés Bello, Avda. Republica 252, Santiago, Chile;
[l]European Southern Observatory, Karl Schwarzschild Strasse 2, D-85748, Garching bei München, Germany;
[m]Queen's University Belfast, Belfast, County Antrim, BT7 1NN, UK;
[n]Incident Angle Oy, Capsiankatu 4 A 29, FI-20320 Turku, Finland;
[o]Centro de Investigaciones en Optica A. C., Loma del Bosque 115, Lomas del Campestre, 37150 Leon Guanajuato, Mexico;
[p]Tel Aviv University, Department of Astrophysics, 69978 Tel Aviv, Israel;
[q]INAF – Osservatorio Astronomico di Arcetri, Largo Enrico Fermi 5, I-50125 Firenze, Italy;
[r]INAF - Istituto di Astrofisica e Planetologia Spaziali, Via Fosso del Cavaliere, I-00133 Roma, Italy;
[s]Tuorla Observatory, Department of Physics and Astronomy, University of Turku, FI-20014
[h]University of Turku, Finland;
[t]Aarhus University, Ny Munkegade 120, D-8000 Aarhus, Denmark.

## ABSTRACT

Son Of X-Shooter (SOXS) is a double-armed (UV-VIS, NIR) spectrograph designed to be mounted at the ESO-NTT in La Silla, now in its Assembly Integration and Verification (AIV) phase. The instrument is designed following a modular approach so that each sub-system can be integrated in parallel before their assembly at system level. INAF-Osservatorio

Astronomico di Padova will deliver the Common Path (CP) sub-system, which represents the backbone of the entire instrument. In this paper, we describe the foreseen operation for the CP alignment and we report some results already achieved, showing that we envisaged the suitable setup and the strategy to meet the opto-mechanical requirements.

**Keywords:** Son of X-Shooter, New Technology Telescope, Spectroscopy, AIV, Astronomical Instrumentation, Transients

# 1. INTRODUCTION

Son Of X-Shooter [1][2][3][4][5] (SOXS) is a double-armed (UV-VIS, NIR) spectrograph designed to be mounted at the Nasmyth focus of the ESO – New Technology Telescope in La Silla. Its main task is the spectroscopic follow-up of transient sources. At the time of writing, the instrument is in its Assembly Integration and Verification (AIV) phase. SOXS is designed following a modular approach: it is composed by a Common Path (CP) [6], the backbone of the instrument, which splits the starlight toward two different spectrographs [7][8][9][10][11][12] (ranging in 0.35-0.85 and 0.8-2.0 μm intervals), a Calibration Unit (CU) [13] and an Acquisition Camera (AC) [14][15]. These sub-systems are going to be integrated in parallel in different consortium premises [16], before their assembly at the system level: INAF-Osservatorio Astronomico di Padova should deliver the CP. The strategy for the CP alignment follows an optomechanical approach: all components are positioned on the CP bench using a mechanical metrological instrument (a Coordinate Measuring Machine). Then, the alignment is fine-tuned and validated with optical feedback with an on-axis collimated source and a telescope simulator. The integration is followed by a series of tests on the PSF quality, on the flexures and their compensation [17], on the functionality of the electronic components [18][19], and on the instrument software [20][21].

In Figure 1, the SOXS CP is depicted with an optical and mechanical [22][22][24] cut in the right and left panels respectively. The F/11 light beam comes from the NTT Nasmyth bending, delivering the focal plane into the CP on a special multi-purpose flat mirror (*Pierced mirror* on *AC slide*, in the picture). This mirror is mounted on a linear stage, which allows for four observing configurations: (i) a slit-like hole on the mirror provides the spectroscopic mode, sending the light toward the instrument: the marginal field light is folded toward the AC, mounted on the top on the CP, for secondary guiding operations. (ii) The mirror has a pinhole used for aligning purposes. (iii) An area of the mirror is not pierced so that all light can be delivered to the AC for photometry. (iv) Finally, next to the mirror, a pellicle beam splitter is mounted to check the mutual alignment of the two spectrographs in the system AIV phase. After the pierced mirror, a dichroic splits the UV-VIS wavelengths (350 – 850 nm) from the NIR ones (800 – 2000 nm) allowing a wavelength overlap of 50 nm for arms cross-calibration. The UV-VIS channel develops with a folding mirror (UVVIS-FM), an Atmospheric Dispersion Corrector (ADC) (actually the ADC prisms are glued to doublets which change the focal number to F/6.5), then a tip-tilt piezo mirror (UVVIS-TT) envisaged to compensate flexures during de-rotation and finally a field lens (UVVIS-FL). The NIR arm has its NIR-FM and NIR-TT, then a doublet (NIR-DL), mounted on a linear stage both for de-focus tuning and to change the F/11 into F/6.5, then a cold stop and a NIR-FL. Besides these components, just after the entrance shutter, a folding mirror is mounted on a linear stage (CU slide and mirror) to switch between the observation channel and the CU (mounted under the CP) input.

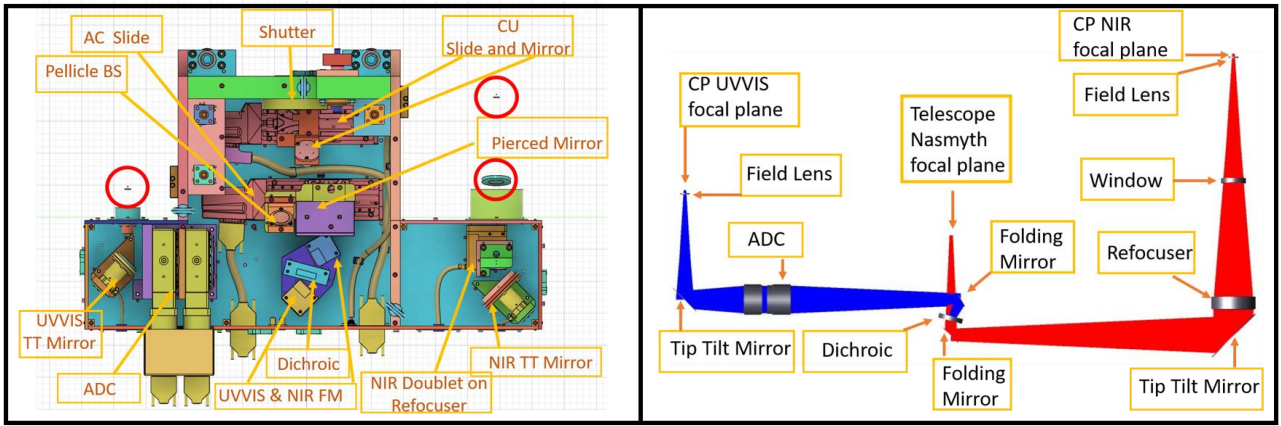

Figure 1. Mechanical (left) and optical (right) sketches of the SOXS Common Path. The optical components (UVVIS field lens, NIR window and NIR field lens) in the red circles are formally a part of the CP, but their mechanical mount should fit the mechanical benches of the spectrographs.

At the time of writing (November 2020), the only missing components are the ADC and the pierced mirror: considering their foreseen delivery dates (Fall 2020), these do not represent a showstopper in the AIV flow.

In this paper, we describe the strategy and the current of the CP AIV, considering the activities which cover the period from the beginning of June to October 2020. Contrary to the expectations claimed in the first abstract delivered for this SPIE proceeding, we are not presenting the results of the acceptance test, since the ongoing COVID19 pandemic affected our schedule (INAF-OA Padova laboratories were open with a restricted regime) and interrupted some providers activities, resulting in a delay of about four months.

## 2. SETUP CONCEPT AND CHARACTERIZATION

Given the relatively small dimensions of the CP (about 710 x 395 x 200 mm, 45 kg), we opt for an optical bench to perform its AIV (we remind the reader that the System AIV will be carried out on a Nasmyth simulator flange see [16]). We fix the CP to the laboratory bench using two of its three kinematic mounts [25], i.e. the actual mechanical interfaces of the CP with the SOXS flange (Figure 2).

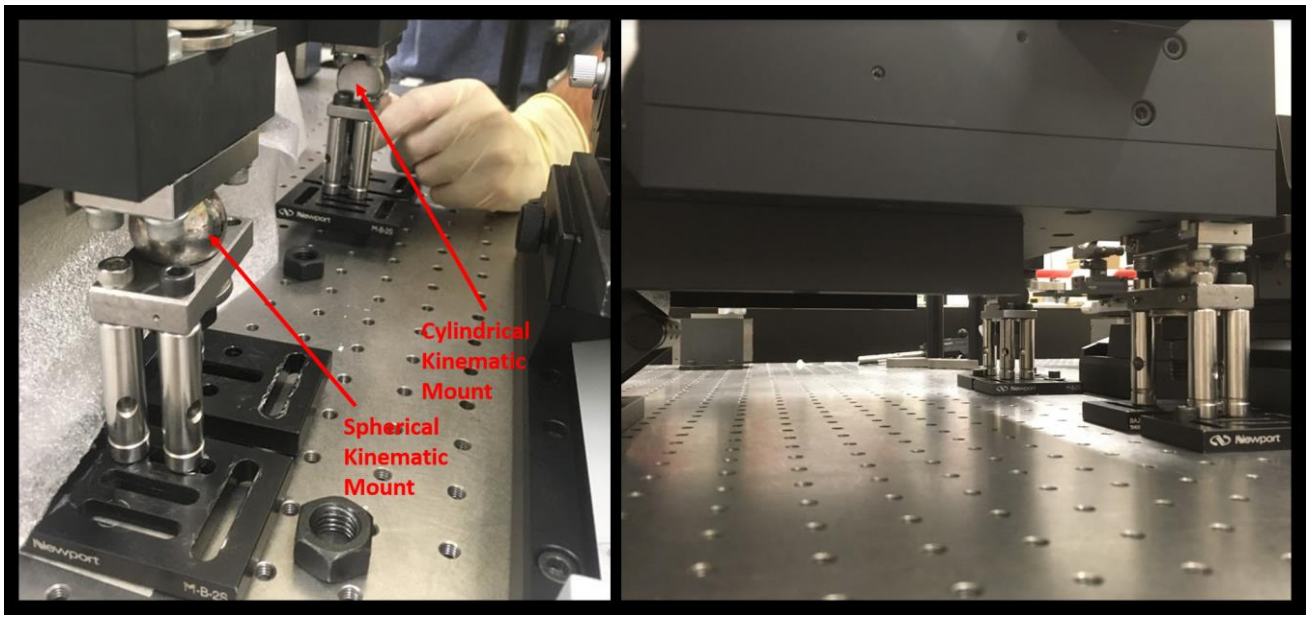

Figure 2. CP fastened to the bench through Kinematic Mounts.

The general approach is opto-mechanical: components are placed in their nominal position with a coordinate measurement machine (CMM) or a portable CMM, then optical feedbacks are defined in order to fine tune the system when the optical precision is higher with respect to the mechanical one. Figure 3 shows the working bench.

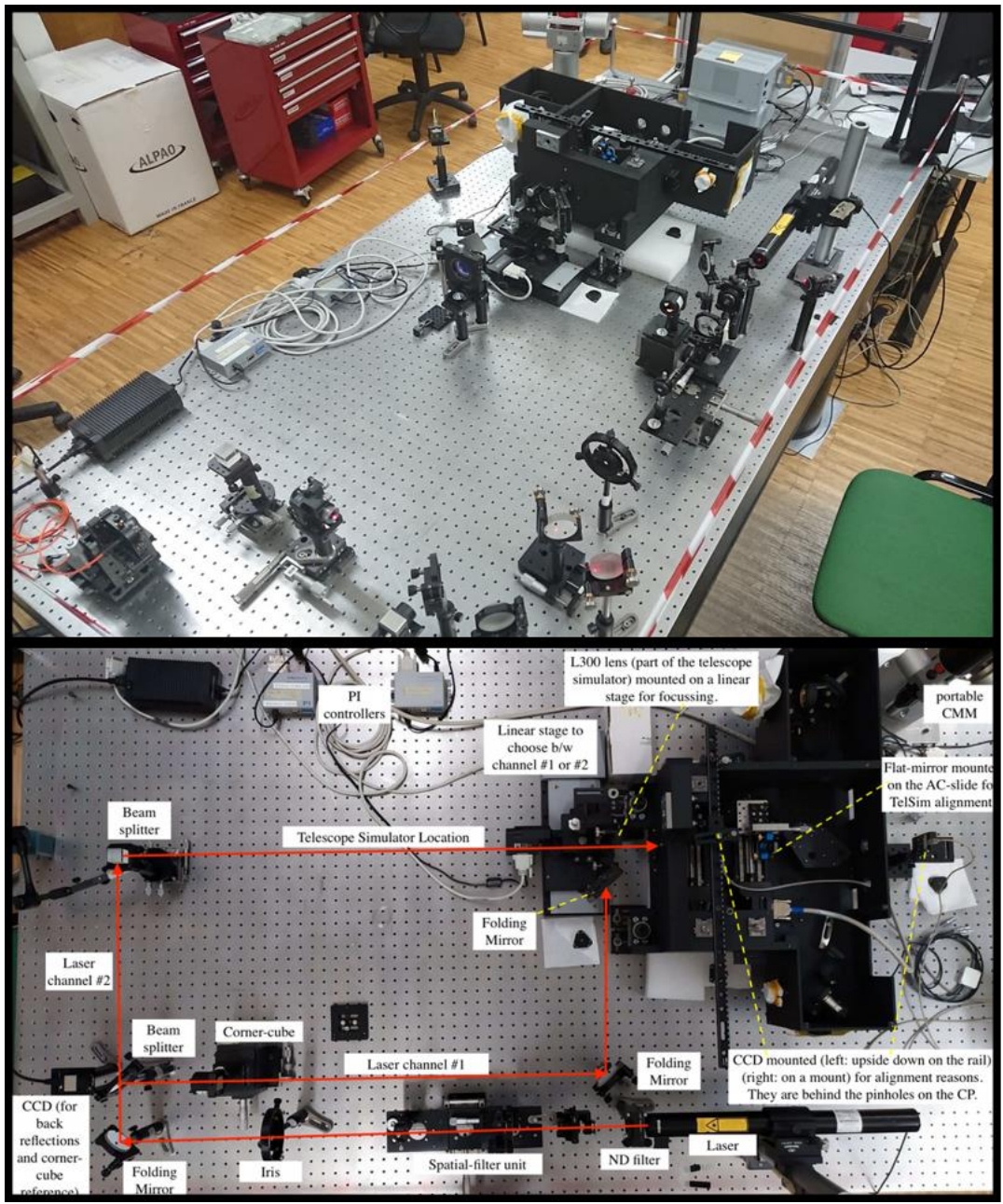

Figure 3. Two different view of the CP working bench. In the bottom panel the two laser channels are represented in red.

We decided to build two co-aligned laser (632.8 nm) channels: channel #1 is used for the on-axis operations, while #2 is used as a reference to align a telescope simulator (Section 3) fed with white (650-1100 nm) light for focus positioning of components with optical power, for off-axis and different wavelengths tests and for the footprint checks. The materialization of the optical axis of the system is constrained by the presence of characterized pinholes (PH) into the mechanical structure (Figure 4) of the CP through which the laser should pass: there are two PHs for the incoming beam, one before the shutter and on the rear part of the bench; two additional lateral PH fix the direction of the 90 deg bent beam; the exit beam will be constrained by placing a characterized CCD (Section 5). All PHs are positioned by means of a CMM considering a reference system defined on the mechanical structure. Laser channel #2 is centered and adjusted in tip-tilt by an iterated procedure of flux maximization, measured by CCDs downstream of the PHs. Laser channel #1 is co-aligned to the #2 by a centroid analysis on the same CCDs, after the removal of the PHs. For the channel #2, the accuracy of the method comes from a combination of possible errors in characterizing the PHs (center position with respect to their mechanical mounts), in mounting the PHs (CMM precision), and finally in the sensitivity of the flux and centroiding methods. We evaluate the combination of these contributes as ±25 µm for the centering and < 10 arcsec for the tilt. Laser channel #1 is co-aligned to the #2 within 14 µm in centering and 3 arcsec in tilt. It is possible to switch between laser channel #1 and #2 by inserting the last folding mirror of the channel #1 with a motorized stage which will host also the last lens of the telescope simulator (Section 3).

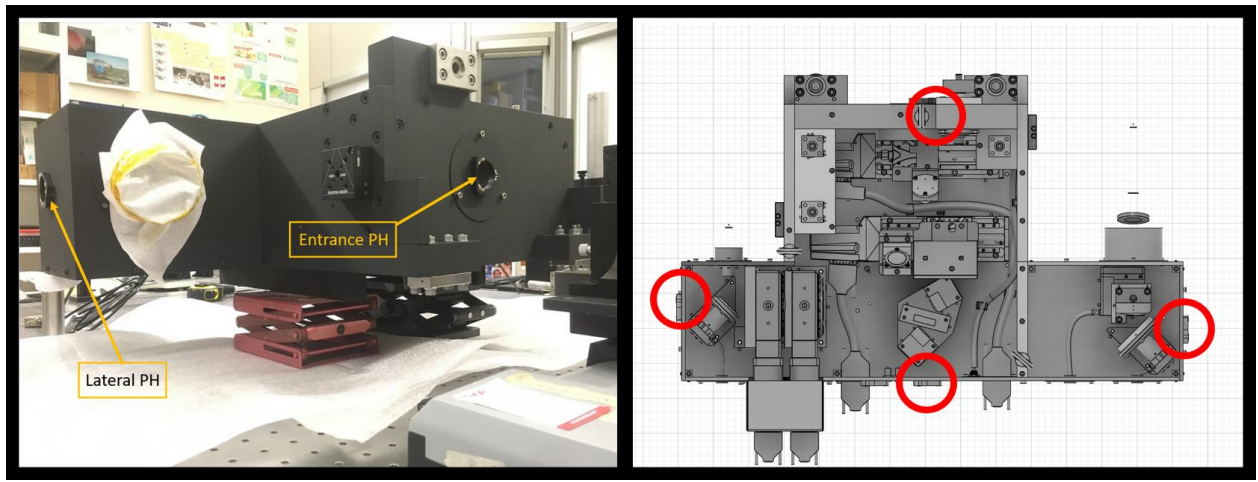

Figure 4. Four pinoles have been characterized and positioned with a CMM on the CP bench. They provide the tool to materialize the optical axis.

## 3. TELESCOPE SIMULATOR

The telescope simulator is designed to match the F/11 of the incoming telescope beam and to create a seeing limited (we consider the value 0.5 arcsec) point spread function. It is composed by a UV-VIS-NIR lamp, a fiber which illuminates an f = 750 mm (hereafter, L750) doublet, followed by a diaphragm and an f = 300 mm (hereafter, L300) lens which refocuses the light in the SOXS entrance focal plane. First, we define one reference for centering the lenses by measuring a centroid on a CCD positioned just after the entrance of the CP: we illuminate the system with laser channel #2 with no additional optics. The centroid reference for the TT is defined on the CCD for back reflection (bottom left in the Figure 3) after the alignment of a corner cube, used to define the auto-collimation path. Then, we place L300 roughly in its nominal distance with respect to the focal plane. We iterate the centering and tip-tilt (centroid from back-reflected light from L300) adjustments by comparing the centroids measurements with and without L300. We center the diaphragm into the L750 mount by shimming it and then we approach L750 alignment with the same steps used for L300. We add a point like source in the laser path and tune its position to provide a collimated beam between the two lenses (checked by a wedge plate). We place the fiber, iterating the lateral positioning and its defocus by comparing the position and dimension of the PSF with the PSF created by the service point like source. Finally, we fine tune the L300 focus by reaching a cat-eye configuration on a flat mirror positioned with the pCMM in the CP focal plane. All the measurements implying a centroid analysis should be considered with a precision of 8 µm, coming from stability checks and from arbitrary parameters in the determination of the centroid.

At the end of the alignment, we reach the values in Table 1 for the positions of the components with respect to the reference axis (laser channel #2):

Table 1. The co-alignment of the Telescope Simulator components with respect to the reference axis (laser channel #2).

| Component | Centering [µm] | Tilt [arcsec] | Defocus [µm] |
|---|---|---|---|
| **L300** | 6 | 2.4 | 5 |
| **L750** | 8 | 2.9 | -- |
| **Fiber** | 5 | -- | 10 |

The PSF captured on the CP focal plane (twenty images average, background subtracted) is shown in Figure 5. The PSF 90% encircled energy diameter matches the predicted one (Figure 6): measured 40 µm vs. theoretical 39 µm.

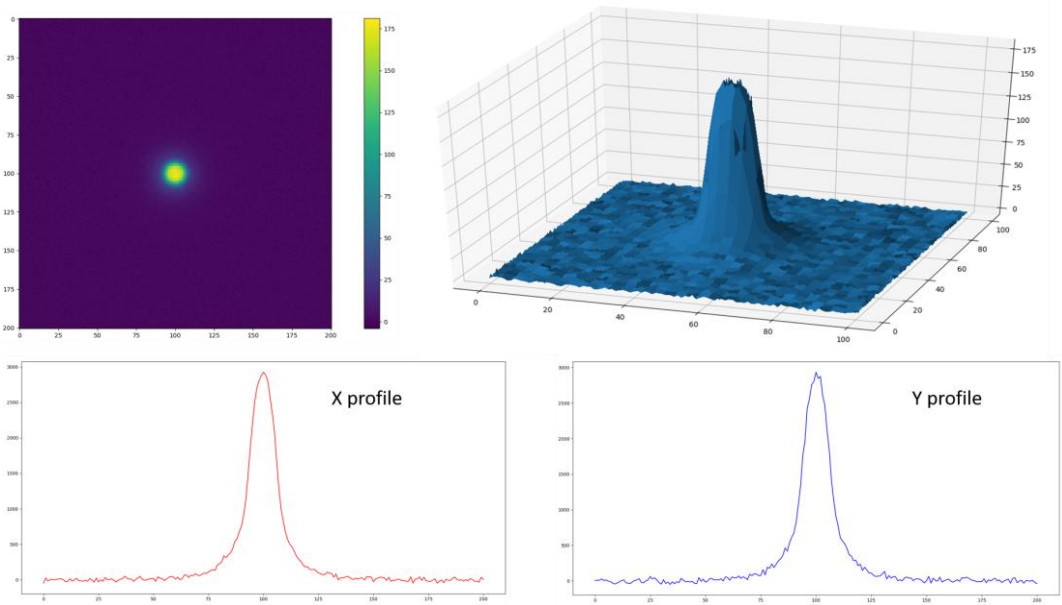

Figure 5. Visualization of the Telescope Simulator PSF on CP focal plane (coordinates in [1.67 µm-px]).

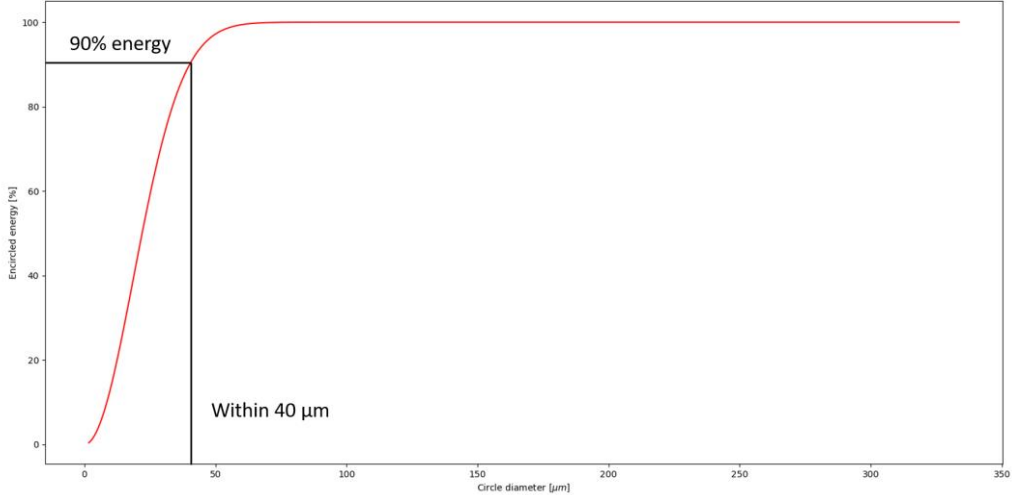

Figure 6. For the resulting PSF of the Telescope Simulator on the CP entrance focal plane, the energy percentage contained in a circle (centered on the PSF centroid) of a certain diameter.

# 4. STRATEGY FOR THE COMMON PATH COMPONENTS

In this Section, we describe the baseline procedure for every element in the CP. Consider that the alignment of such elements will begin last week of November 2020, so results are not provided in this paper. The alignment tolerances for the CP optical components positions are ± 200 µm in centering (except for the ADC: ± 50 µm) and ± 720 arcsec in tip-tilt (except for the NIR-FM and NIR-TT: ± 360 arcsec, and ADC: ± 180 arcsec). We do not describe here the integration of the following components: UVVIS-FL, NIR-FL, NIR-WIN (red circles in Figure 1, left panel) since they are mechanically coupled with the benches of the spectrographs.

## 4.1 Dichroic

We mount the dichroic in nominal position with a pCMM. All corrections are made by shimming (Figure 7): every mount can be shimmed on three lateral point and under the bottom surface.

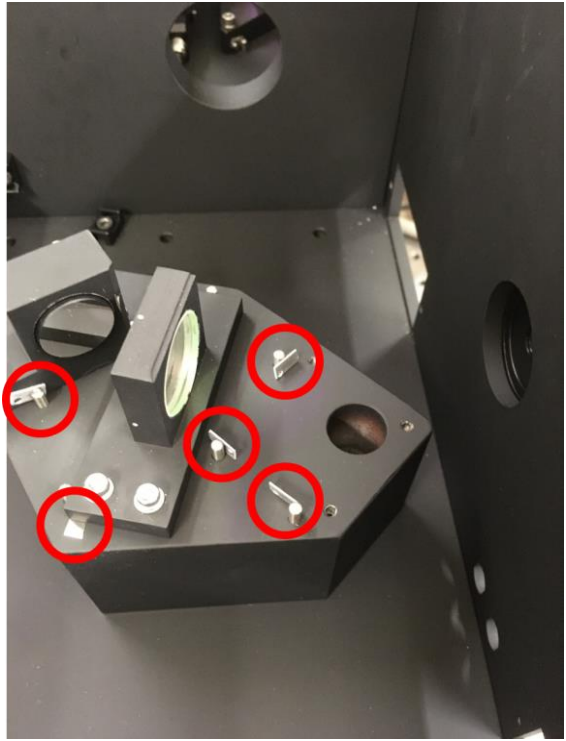

Figure 7. Some of the shimming adjustments for the mechanical block hosting the UVVIS-FM, the dichroic, and the NIR-FM (removed here).

## 4.2 UVVIS-FM and NIR-FM

For both these mirrors, we illuminate the system with laser channel #2. A pentaprism is inserted in the system to provide the 90 deg beam bent. The pentaprism position is defined by maximizing the flux after the lateral PHs. After the removal of the PHs, we take reference centroids. We replace the pentaprism with the actual mirrors: UVVIS-FM and NIR-FM are centered with the pCMM, and their tilts are corrected by comparing the centroids that they provide. All corrections are made by shimming.

## 4.3 UVVIS-TT and NIR-TT

The CP bench can hold a characterized CCD (Section 5) that defines the position of the UV-VIS and NIR arms focal planes. UVVIS-TT and NIR-TT are centered with the pCMM, and their tilts are corrected by targeting the expected pixel on the characterized CCD. All corrections are made by shimming. We remind here that the TT mirror are mounted on tip – tilt piezo stages: their main function is the flexure compensation for different orientation of the instrument, but they can also be used to correct the alignment in case the shimming precision would result not enough. For this reason, it is important to assemble them with their stages at middle range, in order to take advantage of the whole range.

### 4.4 ADC and NIR-DL

The ADC procurement foresees the delivery of an internally aligned system. For alignment purposes, we consider the ADC as a lens and we follow the prescription described in alignment of the simulator lenses for the centering and the TT adjustments (remind that the SOXS ADC is composed by prisms glued on doublets, resulting in a system of two quadruplets mounted on two rotating stages). For the position in focus, we switch the source and introduce the telescope simulator. Then, we perform a through focus of the ADC by minimizing the size of the PSF on the focal plane. All corrections are made by shimming.

The NIR-DL integration follows the same concept used for the ADC.

### 4.5 AC and CU mirrors

Using the telescope simulator and the AC mirror in the PH configuration, it is possible to define its x-y-z position maximizing the flux behind the mirror itself. The TT adjustments are done with the pCMM and checked for the four positions of the AC slide.

The CU mirror is mounted in nominal position just with mechanical (pCMM) feedback. Any optical fine-tuning can be done when the CU will be integrated, shimming the mechanical interface.

### 4.6 Shutter

During the integration of the optics, the shutter is removed (Figure 8). We will mount it back at the end of the alignment.

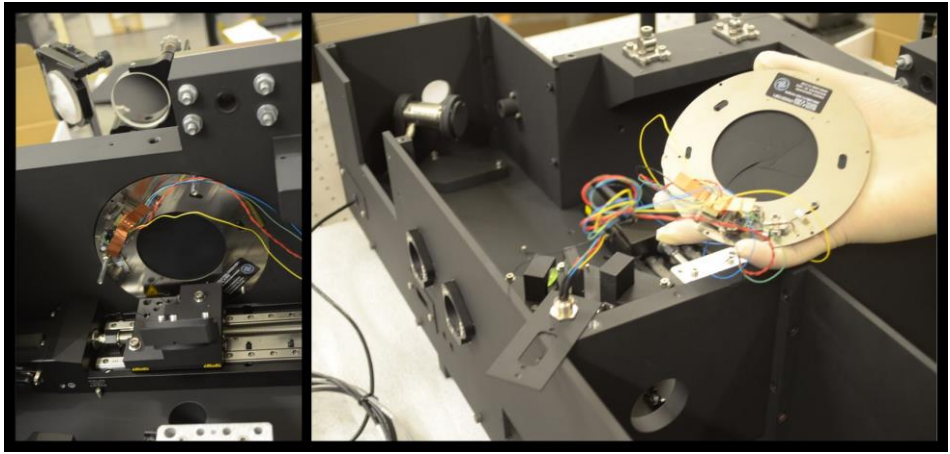

Figure 8. The CP shutter is removed during optics alignment.

## 5. CHARACTERIZATION OF THE SERVICE CCD IN THE FOCAL PLANE

We use a characterized CCD to fix the direction of the exit optical axes bent by the TT mirrors, and to fix the positions along the optical axis of both the ADC and the NIR-DL. The characterization is done with an interferometer and a calibrated metallic sphere, following these steps:

1. We place the metallic sphere in confocal position in front of an interferometer with a spherical element.
2. We measure the center of the sphere with a pCCM.
3. We place a CCD in the focal plane of the interferometer with a through focus analysis.
4. We compute the centroid of the spot on the CCD.

5. We measure the case planes of the CCD with a pCMM.
6. We calculate the geometrical relation between the coordinate system of the illuminated pixel and the center of the sphere.
7. Every time one should be on focus in a fixed point in space, the sphere is used to materialize that point, then the CCD is positioned according to the geometrical relation found. The proper pixel should host the centroid.

## 6. CONCLUSIONS

We described the combined opto-mechanical alignment strategy for the CP of SOXS, which, at the time of writing, is an ongoing process in the INAF-Osservatorio Astronomico di Padova laboratories. We reported the already available results about the alignment of the optical sources used to provide feedbacks for the positioning of the CP components: laser channel #2 materializes the optical axis defined by the PHs mounted in the CP structure within ±25 µm for the centering and < 10 arcsec for the tilt. Laser channel #1, and the telescope simulator components are co-aligned with respect to laser channel #2 with precisions smaller than 15 µm and 3 arcsec. These numbers should be compared with the typical positioning tolerance accepted, since the approach used to prepare the setup is similar in terms of tools, feedbacks and operations to the one for the optical component alignment. Given the precision of the setup, the PSF quality on the CP focal plane provided by the telescope simulator as well as the nominal precision of the mechanical tool used (portable CMM), we can confirm the robustness of the strategy and of the built setup.